\documentclass[aoas,MSNbibl,nameyear,dvips]{arximspdf}
\usepackage{algorithm}
\usepackage{algorithmic}
\usepackage{graphicx}

\doi{10.1214/13-AOAS691} 
\volume{8}
\issue{1}
\pubyear{2014}
\firstpage{377}
\lastpage{405}

\makeatletter

\newcommand{\rright}{\right}
\newcommand{\lleft}{\left}
\newcommand{\Z}{\mathbf{Z}}
\newcommand{\X}{\mathbf{X}}
\newcommand{\A}{\mathbf{A}}

\newcommand{\bu}{\mathbf{u}}
\newcommand{\bv}{\mathbf{v}}
\newcommand{\bgamma}{\bolds{\gamma}}
\newcommand{\bPi}{\bolds{\Pi}}
\newcommand{\balpha}{\bolds{\alpha}}
\newcommand{\btau}{\bolds{\tau}}
\newcommand{\btheta}{\bolds{\theta}}
\newtheorem{prop}{Proposition}[section]
\makeatother

\begin{document}
\begin{frontmatter}

\title{The random subgraph model for the analysis of an ecclesiastical
network in Merovingian Gaul}
\runtitle{The random subgraph model}

\begin{aug}
\author[A]{\fnms{Yacine} \snm{Jernite}\thanksref{m1,m4}\ead[label=e1]{yacine.jernite@polytechnique.edu}},
\author[B]{\fnms{Pierre} \snm{Latouche}\corref{}\thanksref{m1}\ead[label=e2]{pierre.latouche@univ-paris1.fr}},
\author[C]{\fnms{Charles} \snm{Bouveyron}\thanksref{m2}\ead[label=e3]{charles.bouveyron@univ-paris1.fr}},
\author[D]{\fnms{Patrick} \snm{Rivera}\thanksref{m3}\ead[label=e]{patrick.rivera@univ-paris1.fr}},
\author[D]{\fnms{Laurent} \snm{Jegou}\thanksref{m3}\ead[label=e]{laurent.jegou@univ-paris1.fr}}
\and
\author[D]{\fnms{St\'ephane} \snm{Lamass\'e}\thanksref{m3}\ead[label=e]{stephane.lamasse@univ-paris1.fr}}
\runauthor{Y. Jernite et al.}
\affiliation{Laboratoire SAMM, EA 4543, Universit\'e Paris 1 Panth\'eon--Sorbonne\thanksmark{m1},
Laboratoire MAP5, UMR CNRS 8145, Universit\'e Paris Descartes and\\ Sorbonne Paris Cit\'e\thanksmark{m2}
Laboratoire LAMOP, UMR 8589, Universit\'e Paris 1 Panth\'eon--Sorbonne\thanksmark{m3}
and Ecole Polytechnique\thanksmark{m4}}
\address[A]{Y. Jernite\\
Laboratoire SAMM, EA 4543\\
Universit\'e Paris 1 Panth\'eon-Sorbonne\\
90 rue de Tolbiac\\
75634 Paris Cedex 13\\
and\\
Ecole Polytechnique\\
91128 Palaiseau Cedex\\
France} 
\address[B]{P. Latouche\\
Laboratoire SAMM, EA 4543\\
Universit\'e Paris 1 Panth\'eon-Sorbonne\\
90 rue de Tolbiac\\
75634 Paris Cedex 13\\
France\\
\printead{e2}}
\address[C]{C. Bouveyron\\
Laboratoire MAP5, UMR CNRS 8145\\
Universit\'e Paris Descartes\\
\quad and Sorbonne Paris Cit\'e\\
45 rue des Saints P\`eres\\
75006 Paris\\
France}
\address[D]{P. Rivera\\
L. Jegou\\
S. Lamass\'e\\
Laboratoire LAMOP, UMR 8589\\
Universit\'e Paris 1 Panth\'eon-Sorbonne\\
17 rue de la Sorbonne\\
75005 Paris\\
France}
\end{aug}

\received{\smonth{12} \syear{2012}}
\revised{\smonth{10} \syear{2013}}

%
\begin{abstract}
In the last two decades many random graph models have been proposed to
extract knowledge from
networks. Most of them look for communities
or, more generally, clusters of vertices with homogeneous connection
profiles. While the first models focused on networks with binary
edges only, extensions now allow to deal with
valued networks. Recently, new models were also introduced in order to
characterize connection patterns in networks
through mixed memberships. This work was motivated by the need of
analyzing a historical network where a partition of the vertices is
given and where edges are typed. A
known partition is seen as a decomposition of a network into subgraphs
that we propose
to model using a stochastic model with unknown latent clusters. Each
subgraph has its own mixing vector and sees its vertices associated to
the clusters. The vertices then connect
with a probability depending on the subgraphs only, while the types of
edges are assumed to be sampled from the latent clusters. A variational
Bayes expectation-maximization algorithm is proposed for inference as
well as a model selection criterion for the estimation of the cluster
number. Experiments are carried out on
simulated data to assess the approach. The proposed methodology is
then applied to an ecclesiastical network in Merovingian Gaul. An R
code, called \emph{Rambo}, implementing the inference algorithm is
available from the
authors upon request.
\end{abstract}

%

\end{frontmatter}

\section{Introduction}\label{sec1}

Since the original work of \citet{moreno1934shall} on sociograms, network
data has become ubiquitous in Biology [\citet{ArticleAlbert2002,ArticleMilo2002,ArticlePalla2005}] and
computational social sciences
[\citet{snijders1997estimation}]. Applications range from the study of
gene regulation processes to that of social interactions.
Network analysis was also applied recently to a medieval social network
in \citet{villa2008mining}, where the authors find a~clustering of
vertices through kernel methods. Both deterministic and probabilistic methods
have been used to seek structure in these networks, depending on prior
knowledge and assumptions on
the form of the data. For example, \citet{hofman2008bayesian}
looked for a partition of the vertices where the clusters exhibit a
transitivity property.
The model of \citet{handcock2007model}, on the other hand, assumes the
relations to be conditioned on the projection of the vertices in a
latent social space.
Notable among the community discovery methods, though asymptotically
biased [\citet{bickel2009nonparametric}], are
those based on the modularity score given by
\citet{girvan2002community}.

Many of the other currently used methods derive from the stochastic block
model (SBM) [\citet{wang1987,nowicki2001estimation}], which is a probabilistic
generalization [\citet{fienberg1981categorical}] of the method applied by
\citet{white1976social} to Sampson's famous monastery data. SBM assumes that
each vertex belongs to a hidden cluster and that connection probabilities
between a pair of vertices depend exclusively on their clusters, as in
\citet{frank1982cluster}.
The parameters and clusters are then inferred to optimize a criterion,
usually a lower
bound of an integrated log-likelihood. Thus, \citet{latouche2009overlapping} used
an approximation of the marginal log-likelihood, while \citet{daudin2008mixture}
considered a~Laplace approximation of the integrated classification
log-likelihood. A nonparametric Bayesian approach was also introduced
by \citet{kemp2006learning} to estimate the number of clusters while
clustering the vertices.
SBM was extended by the mixed membership stochastic block model (MMSBM)
[\citet{airoldi2008mixed}], which allows a vertex to belong to
different clusters in its relations toward different vertices, and by
the overlapping
stochastic block model (OSBM) [\citet{latouche2009overlapping}], which
allows a
vertex to belong to no cluster or to several at the same time. More
recent works
focused on extending MMSBM to dynamic networks [\citet{xing2010state}]
or dealing with nonbinary networks, such as networks with weighted edges
[\citet{soufiani2012graphlet}]. \citet{goldenberg2010survey} and
\citet{salter2012} provide
extensive reviews of statistical network models.

In this paper we aim at clustering the vertices of networks with
typed edges and for which a partition of the nodes into subgraphs is observed
and bears some
importance in their behavior. For example, one may be interested in
looking for latent clusters in a worldwide social
network describing social interactions between individuals where
different countries, or at
a different scale, different regions of the world, have
different connectivity patterns. We might
also observe the same kind of phenomenon between different
scientific fields in a citation network. This kind of network may be
modelled using generalized linear models
[\citet{fienberg1981categorical}] by incorporating the observed
partition information as covariates and the clusters serving as
random effects or a $p1$ model
[\citet{holland1981}], where the
clusters allow for the estimation of interactions. However, we
consider here a different strategy and propose an extension of the
SBM model which has the advantage of relying on easy to interpret
parameters. Indeed, SBM parameters are not expressed through nonlinear
functions like the log or logistic functions and this allows
an easy interpretation for nonstatisticians.

This point is of crucial
interest in this work because we aim at providing historians with
insight into the relationships between
ecclesiastics and notable people in the kingdoms that made up
Merovingian Gaul,
by analyzing a network characterizing their different kinds of
relations. Specifically, the data set focuses on the relationships
between individuals built
during the ecclesiastical councils which took place in Gaul during the 6th
century. These councils were convened under the
authority of a bishop to discuss specific questions relating to the Church.
Though consisting mainly of clergymen, laics would also occasionally take
part, as representatives of the secular power or experts in the questions
discussed. These assemblies shaped a significant part of that period, and
we are interested in discovering how they reflected the relationships between
various groups of individuals. For this network, extra information
on the vertices, namely, a geographical partition, is available,
associating each
individual to a specific kingdom. This partition induces a
decomposition of the network into subgraphs and we aim at modelling
the connection pattern of each subgraph through latent clusters.

Thus, in this paper, we propose a new model, that we call the random subgraph
model (RSM), for the analysis of directed networks with typed edges
for which a~partition of the vertices is available. On the one hand,
we consider that
the probability of observing an edge between two vertices depends
solely on the subgraphs to which the vertices belong. On the other
hand, we assume that each vertex belongs to
a hidden cluster, with a probability depending on its subgraph. Then,
if a relation is present, its type is drawn from a multinomial distribution
whose parameters depend on the clusters to which the vertices
belong. Let us emphasize that the
latter property allows, once the inference is done, to compare the different
subgraphs.

The choice of proposing a probabilistic rather than a deterministic
model is again motivated by the nature of the historical network we
consider. Indeed, as mentioned in Section~\ref{sececclNet}, the
data set was built from a collection of data at hand using sources
such as council acts or narrative texts. However, the rarity of the
sources only allowed an incomplete or approximate characterization
of the relations between individuals. Therefore, we rely
on the probabilistic framework in order to deal with the uncertainty
on the
edges. Moreover, we emphasize that probabilistic methods for network
analysis are appealing in general because they have been shown to be flexible
and capable of retrieving complex heterogeneous structures in
networks [see, for instance, \citet{airoldi2008mixed,goldenberg2010survey}].

The article is organized as follows. The random subgraph model is presented
along with its inference algorithm in Section~\ref{sec2}, then tested on
simulated data
and compared to other models in Section~\ref{sec3}. Our model is then applied
to the
ecclesiastical network and the results are analyzed from the
historical point of view in Section~\ref{sec4}. Concluding
remarks and possible extensions are finally discussed in Section~\ref{sec5}.

\section{The random subgraph model}\label{sec2}

We consider a directed graph $\mathcal{G}$ with $N$ vertices
represented by its
$N \times N$ adjacency matrix $\mathbf X$ along with a known
partition $\mathcal{P}$ of the vertices into $S$ classes. Our goal
is to cluster the network into $K$ groups with homogeneous
connection profiles, {that is}, estimating a binary matrix $\Z$
such that $Z_{ik}=1$ if vertex $i$ belongs to cluster $k$, $0$
otherwise.

Let us now detail the notation. Each edge $X_{ij}$,
describing the relation between the vertices $i$ and $j$, is typed,
{that is},
takes its values in a finite set $ \{0,\ldots,C\}$. Note that
$X_{ij}=0$ corresponds to the absence of
an edge. We assume that $\mathcal{G}$ does not have any self loop and,
therefore, the entries $X_{ii}$ will not be taken into account. In
order to simplify the
notation when describing the model, we also consider the binary
matrix $\mathbf A$ with entries $A_{ij}$ such that $A_{i,j}=1 \iff
X_{i,j} \ne0$.

We also emphasize that the observed partition $\mathcal{P}$ induces a
decomposition of
the graph into subgraphs where each class of vertices corresponds to a
specific subgraph. We introduce the variable $s_{i}$ which takes its
values in
$\{1,\ldots,S\}$ and is used to indicate in which of the subgraphs
vertex $i$ belongs, for $i\in\{1,\ldots,N\}$.

\subsection{The probabilistic model}\label{sec2.1}

The data is assumed to be generated in three steps. First, the
presence of an edge from vertex $i$ to vertex $j$ is supposed to
follow a Bernouilli
distribution whose parameter depends on the subgraphs $s_i$ and
$s_j$ only:
\[
A_{i,j} \sim\mathcal{B}(\gamma_{s_i,s_j}).
\]
Each vertex $i$ is then associated to a latent cluster with
a probability depending on~$s_{i}$. In practice, if we assume for
now that the number $K$ of latent clusters is known, the variable
$\mathbf Z_{i}$ is drawn from a multinomial distribution:
\[
\mathbf Z_i \sim\mathcal{M}(1;\bolds\alpha_{s_i}),
\]
where
\[
\forall s \in{1,\ldots,S}\qquad\sum_{k=1}^K{\alpha_{s k}}=1.
\]
A notable point of the
model is that we allow each subgraph to have different mixing
proportions $\balpha_{s}$ for the latent clusters. We denote
hereafter $\balpha=(\balpha_{1},\ldots,\balpha_{S})$. Finally, if an
edge between $i$
and $j$ is present, {that is}, $A_{ij}=1$, its type $X_{ij}$ is sampled
from a multinomial distribution with parameters depending on the
latent clusters. Thus, if $i$ belongs to cluster $k$ and $j$ to
cluster $l$,
\[
X_{i,j} | Z_{ik}Z_{jl}=1,\qquad A_{ij}=1 \sim
\mathcal{M}(1, \bolds\Pi_{kl}),
\]
where
the sum over the $C$ types of each vector $\bolds\Pi_{kl}=(\Pi
_{kl1},\ldots,\Pi_{klC})$ is
\[
\forall(k,l) \in \{1,\ldots,K\}^2\qquad \sum_{c=1}^C{
\Pi_{klc}}=1.
\]
If there is no edge between the two vertices, the entry $X_{ij}$ is
simply set
to $X_{ij}=A_{ij}=0$.

The model is therefore defined through the joint distribution
\begin{eqnarray*}
p(\X,\A,\Z|\balpha,\bgamma,\bPi)&=&p(\X,\A|\Z,\bgamma,\bPi)p(\Z|\balpha)
\\
&=& p(\X|\A,\Z,\bPi)p(\A|\bgamma)p(\Z|\balpha),
\end{eqnarray*}
where
\[
p(\X|\A,\Z,\bPi)=\prod_{k,l}^{K}\prod
_{c=1}^{C}\bigl(\Pi_{kl}^{c}
\bigr)^{\sum_{i
\neq j}^{N}\delta(X_{ij}=c)A_{ij}Z_{ik}Z_{jl}}
\]
and
\[
p(\A|\bgamma)=\prod_{i\neq j}^{N}
\gamma_{r_{i},r_{j}}^{A_{ij}}(1-\gamma _{r_{i},r_{j}})^{1-A_{ij}}.
\]
Finally,
\[
p(\Z|\balpha)= \prod_{i=1}^{N}\prod
_{k=1}^{K}\alpha_{r_{i},k}^{Z_{ik}}.
\]
We refer to the \hyperref[app]{Appendix} for the detailed calculation of the complete
data log-likelihood associated to the RSM model and summarize the
model parameters in Table~\ref{tabnotations}.

%
\begin{table}[b]
\caption{Summary of the notation used in the paper}\label{tabnotations}
\begin{tabular*}{\textwidth}{@{\extracolsep{\fill}}ll@{}}
\hline
\textbf{Notation} & \textbf{Description}\\
\hline
$\X$ & Adjacency matrix. $X_{ij}\in\{0,\ldots,C\}$ indicates the edge type\\
$\A$ & Binary matrix. $A_{ij}=1$ indicates the presence of an edge\\
$\Z$ & Binary matrix. $Z_{ik}=1$ indicates that $i$ belongs to cluster $k$ \\
$N$ & Number of vertices in the network\\
$K$ & Number of latent clusters \\
$S$ & Number of subgraphs \\
$C$ & Number of edge types \\
$\balpha$ & $\alpha_{sk}$ is the proportion of cluster $k$ in subgraph $s$ \\
$\bPi$ & $\Pi_{klc}$ is the probability of having an edge of type $c$ between vertices of clusters $k$ and $l$ \\
$\bgamma$ & $\gamma_{rs}$ probability of having an edge between
vertices of subgraphs $r$ and $s$ \\
\hline
\end{tabular*}
\end{table}
%
\begin{figure}[b]
\includegraphics{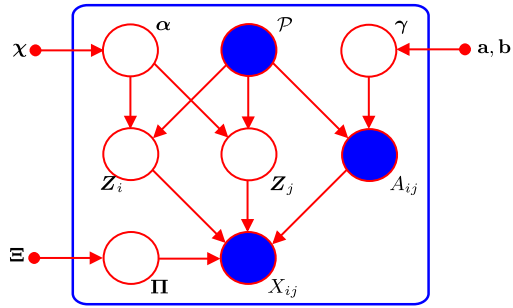}
\caption{Graphical representation of the RSM model.}\label{graphicalmodel}
\end{figure}

We point out that the choice of separating the role of the known
subgraphs and the latent clusters was motivated by historical
assumptions on the creation of relationships between individuals in
Gaul during the $6$th century. These assumptions were
at the core of the study of the ecclesiastical
network we \mbox{consider} in this paper. An alternative approach would
consist in allowing the presence of an edge and its type to depend on
both the subgraphs and latent clusters. However, this would
dramatically increase the number of model parameters to
be estimated. Indeed, for a network with $S=6$, $K=6$, and $C=4$, it would
require $K^{2}S^{2}(C+1) + SK = 6516$ parameters while RSM only
involves $S^{2} + K^{2}C + SK = 216$ parameters.

\subsection{Bayesian framework}\label{sec2.2}

We consider a Bayesian framework and introduce conjugate
prior distributions. Thus, since $\Z_{i}$ is sampled from a multinomial
distribution, we rely on a Dirichlet prior to model the
parameters $\balpha_{s}$:
\[
p(\balpha_{s})=\operatorname{Dir} \bigl(\balpha_{s};
\chi_{s1}^{0},\ldots,\chi _{sK}^{0}
\bigr)\qquad \forall s\in\{1,\ldots,S\}.
\]
A similar distribution is used as a prior distribution for the
parameters $\bPi_{kl}$:
\[
p(\bPi_{kl})=\operatorname{Dir} \bigl(\bPi_{kl};
\Xi_{kl1}^{0},\ldots,\Xi _{klC}^{0} \bigr)\qquad
\forall (k,l)\in\{1,\ldots,K\}^{2}.
\]
If no prior information is available, a common choice in the
literature consists in fixing the hyperparameters of the Dirichlet to
$1/2$, {that is}, $\chi_{sk}^{0}=1/2,\forall (s,k)$ and
$\Xi_{klc}=1/2,\forall(k,l,c)$. Such a distribution corresponds to a
noninformative Jeffreys prior distribution which is known to be
proper [\citet{proceedingsjeffreys1946}]. A uniform distribution can
also be obtained by setting the hyperparameters to $1$.

Finally, since the presence or absence of an edge between a pair of
vertices is
drawn from a Bernoulli distribution, we rely on a beta prior for the
parameters $\gamma_{rs}$:
\[
p(\gamma_{rs})=\operatorname{Beta}\bigl(\gamma_{rs};
a_{rs}^{0},b_{rs}^{0}\bigr)\qquad
\forall(r,s)\in\{1,\ldots,S\}^{2}.
\]
Again, if no prior information is available, both
hyperparameters $a_{rs}^{0}$ and $b_{rs}^{0}$ can be set to $1/2$ or
$1$ to obtain noninformative prior distributions, respectively, a
Jeffreys or a uniform distribution.
Figure~\ref{graphicalmodel} presents the graphical model associated with
the RSM model.

\subsection{Inference with the variational Bayes EM algorithm}\label{sec2.3}\label{ssectionVBEM}

Given the observed~matrices $\mathbf X$ and $\A$, we aim at estimating the
posterior distribution\break $p(\mathbf Z,\balpha,\bgamma,\bPi|\mathbf X,\A
)$, which in turn will allow us to compute
a maximum a posteriori estimate of the clustering structure $\Z$ as
well as the parameters $(\balpha,\bgamma,\bPi)$. Because
this distribution is not tractable, approximate inference procedures
are required. The Markov chain Monte Carlo (MCMC) sampling scheme is
a widely used approach which consists in sampling from tractable
conditional distributions. After a burn-in period, samples are assumed
to be drawn from the true posterior
distribution. One of the main advantages of the MCMC algorithm is that it
can characterize the uncertainty in model
parameters. Moreover, the convergence of the Markov chain and
therefore the quality of the approximation can be tested.

Unfortunately, the MCMC algorithm has a poor scaling with sample sizes. This
motivated the work of \citet{daudin2008mixture} who proposed a
variational approach for the SBM model which can deal with large
networks contrary to the MCMC method of
\citet{nowicki2001estimation}. In general, the main drawback of
variational techniques is that, although they can produce a good
estimate of the model parameters or find the mode of the posterior
distribution, they usually cannot uncover the uncertainty in the
model parameters and tend to underestimate posterior variances.
Furthermore, the quality of the variational
approximation cannot be tested in most cases since the KL divergence
between the true and approximate posterior distribution is not
tractable.

However, recent results [\citet{celisse2012,mariadassou2013}] gave some
new insights on the form
of the true posterior distribution in the case of the SBM model and
showed that the corresponding variational estimates were
consistent. In light of these recent results and because we aim at
proposing an inference procedure capable of handling large networks,
we rely in the following on a variational Bayes EM (VBEM) algorithm.

Thus, given a
distribution $q(\Z,\balpha,\bgamma,\bPi)$, the marginal log-likelihood
can be computed in two terms,
\[
\log p(\X,\A)=\mathcal{L}(q)+\mathrm{KL}\bigl(q(\cdot)\|p(\cdot|\X,\A)\bigr),
\]
where
$\mathcal{L}$ is defined as follows:
\[
\mathcal{L}(q)=\sum_{\Z}{\int_{\balpha,\bgamma,\bPi}{
q(\Z,\balpha,\bgamma,\bPi)\log\biggl(\frac{p(\X,\A,\Z,\balpha,\bgamma,\bPi
)}{q(\Z,\balpha,\bgamma,\bPi)}\biggr) \,d\balpha \,d\bgamma \,d
\bPi}}
\]
and the KL divergence is given by
\begin{eqnarray*}
&& \mathrm{KL}\bigl(q(\cdot)\|p(\cdot|\X)\bigr)
\\
&&\qquad =-\sum_{\Z}{\int
_{\balpha,\bgamma,\bPi}{ q(\Z,\balpha,\bgamma,\bPi)\log\biggl(\frac{p(\mathbf
Z,\balpha,\bgamma,\bPi|\X, \A)}{q(\mathbf Z,\balpha,\bgamma,\bPi)}
\biggr) \,d\balpha \,d\bgamma \,d\bPi}}.
\end{eqnarray*}
Finding the best approximation of the posterior distribution $p(\Z,\balpha,\bgamma,\bPi|\X,\A)$ in the
sense of the KL divergence becomes equivalent to finding $q(\cdot)$
that maximizes
the lower bound $\mathcal{L}(q)$ of the integrated log-likelihood. To obtain
a tractable algorithm, we assume that $q(\cdot)$ can be fully factorized,
that is,
\[
q(\mathbf Z,\balpha,\bgamma,\bPi)= \Biggl(\prod_{i=1}^{N}{q(
\mathbf Z_i)} \Biggr) \Biggl(\prod_{s=1}^{S}{
q(\bolds\alpha_s)\prod_{t=1}^{S}{
q(\gamma_{s,t})} } \Biggr) \prod_{k,l}^{K}{q(
\bolds\Pi_{k,l})}.
\]
The functional optimization of the lower bound with respect to
$q(\cdot)$ is performed using a VBEM algorithm (see Algorithm \ref{algoVBEM}). All the optimization equations are given in the
\hyperref[app]{Appendix}. We emphasize that the functional form of the prior
distributions is preserved through the optimization. In
particular, $q(\mathbf{Z})$ is given by
\[
q(\mathbf{Z}) = \prod_{i=1}^{N}q(
\mathbf{Z}_{i})=\prod_{i=1}^{N}
\mathcal {M}(\mathbf{Z}_{i};1,\bolds{\tau}_{i}),
\]
where $\tau_{ik}$ is a variational parameter denoting the probability of
node $i$ to belong to cluster $k$. The approximate posterior
distributions over the other model parameters $(\balpha,\bgamma,
\bPi)$ depend on parameters that we denote
$\btheta= \{\bolds{\chi}, (\mathbf{a},\mathbf{b}),
\bolds{\Xi} \}$, respectively.

\begin{algorithm}[t]
\caption{VBEM algorithm for the RSM model (see text for details)}
\begin{algorithmic}
\STATE Initialize matrix $\btau=\Z$ with k-means
\STATE Initialize hyperparameters $\btheta^{0}= \{\bolds{\chi}^{0},
(\mathbf{a}^{0},\mathbf{b}^{0}),
\bolds{\Xi}^{0} \}$
\STATE Compute $\mathcal{L}(q)$
\WHILE{$|\btheta^{\mathrm{new}}-\btheta^{\mathrm{old}}| \ge\varepsilon$}
\STATE E step: update $\btau$
\STATE M step: update $\btheta^{\mathrm{new}}= \{\bolds{\chi}, (\mathbf
{a},\mathbf{b}),
\bolds{\Xi} \}$
\STATE Compute $\mathcal{L}$
\ENDWHILE
\end{algorithmic}
\label{algoVBEM}
\end{algorithm}

\subsection{Initialization}\label{sec2.4}

The VBEM algorithm, though useful in approximating posterior
distributions of graphical models,
is only guaranteed to converge to a~local optimum [\citet{bilmes1998gentle}].
Strategies to tackle this issue include simulated annealing and the use of
multiple initializations [\citet{biernacki2001strategies}]. In this
work, we choose
the latter option. In order to have a better chance of reaching a global
optimum, VBEM is run for several initializations of a k-means like
algorithm with the following distance $d(i,j)$ between the vertices
$i$ and $j$:
\begin{equation}
\label{eqdist} d(i,j)=\sum_{h=1}^{N}
\delta(X_{ih}\neq X_{jh})A_{ih}A_{jh} +
\sum_{h=1}^{N}\delta(X_{hi}\neq
X_{hj})A_{hi}A_{hj}.
\end{equation}
The first term looks at all possible edges from $i$ and $j$ toward a
third vertex $h$. If both $i$ and $j$ are connected to $h$,
{that is}, $A_{ih}A_{jh}=1$, the edge types $X_{ih}$ and $X_{jh}$ are
compared. By symmetry, the second term looks at all possible edges
from a vertex $h$ to both $i$ as well as $j$, and compare their types.
Thus, the distance computes the
number of discordances in the way both $i$ and $j$ connect to other
vertices or
vertices connect to them. The algorithm starts by sampling the cluster
centers
among all the vertices of the network. It then iterates a two-step
procedure until convergence of the cluster centers. In the first step,
the vertices
are classified into the cluster with the closest center. Each
cluster center is then associated to a vertex minimizing its distance
with all the vertices of the corresponding
cluster.

\subsection{Choice of $K$}\label{sec2.5}

So far, the number $K$ of latent clusters has been assumed to be
known. Given $K$, we showed in Section~\ref{ssectionVBEM} how an
approximation of the posterior distribution over the latent
structure and model parameters could be obtained. We now address the
problem of estimating the number of clusters directly from the data.
Given a set of
values of $K$, we aim at selecting $K^{*}$ for which the marginal
log-likelihood $\log p(\X|K)$ is maximized. However, because
this integrated log-likelihood involves a marginalization over all the
model parameters and latent variables, it is not
tractable. Therefore, we propose to replace the marginal log-likelihood
with its variational approximation, as in
\citet{bookbishop2006,inbooklatouche2009,articlelatouche2012}. Thus,
for each value
of $K$ considered, the VBEM algorithm is applied. We recall that the
maximization of the lower bound induces a minimization of the KL
divergence. After convergence of the algorithm, the lower bound is
used as an approximation of $\log p(\X|K)$ and $K$ is
chosen such that the lower bound is maximized. We prove in the
\hyperref[app]{Appendix} that, if computed right after
the M step of the variational Bayes EM, the lower bound has the following
expression:
\begin{eqnarray*}
\mathcal{L}(q)&=&\sum_{r,s}^{S}\log\biggl(
\frac
{B(a_{rs},b_{rs})}{B(a_{rs}^{0},b_{rs}^{0})}\biggr) +\sum_{s=1}^{S}\log
\biggl(\frac{C(\bolds \chi_{s})}{C(\bolds
\chi_{s}^{0})}\biggr)
\\
&&{}  + \sum_{k,l}^{K} \log\biggl(\frac{C(\bolds
\Xi_{kl})}{C(\bolds\Xi_{kl}^{0})}\biggr)-\sum_{i=1}^{N}
\sum_{k=1}^{K}\tau_{ik}\log(
\tau_{ik}),
\end{eqnarray*}
where
$C(x)=\frac{\prod_{d=1}^{D}\Gamma(x_{d})}{\Gamma(\sum_{d=1}^{D}x_{d})}$
if $x\in\mathbb{R}^{D}$,
$B(a,b)=\frac{\Gamma(a)\Gamma(b)}{\Gamma(a+b)},\forall
(a,b)\in\mathbb{R}^{2}$, and $\Gamma(\cdot)$ is the gamma function. See
the \hyperref[app]{Appendix} for the definition of $a_{rs}$, $b_{rs}$, $\bolds{\chi_{s}}$,
$\bolds{\Xi_{kl}}$ and~$\tau_{ik}$.
%
\begin{figure}[t]
\includegraphics{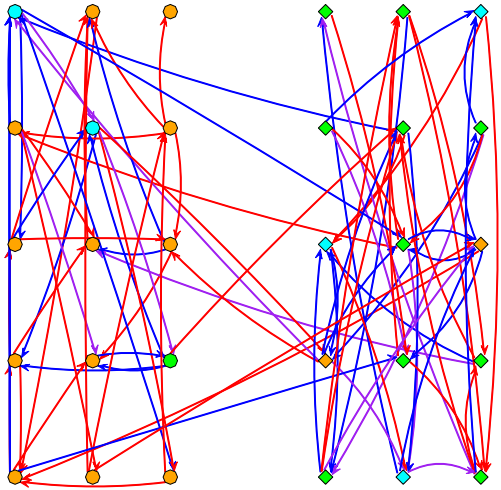}
\caption{Example of a RSM network for $S=2$ subgraphs (indicated by the
node forms), $C=3$ types of edges (indicated by the edge colors) and
$K=3$ clusters to identify (indicated by the node colors).}
\label{example}
\end{figure}

\section{Numerical experiments and comparisons}\label{sec3}

In this section we first run experiments aimed at proving the
validity of our model, focusing on the ability of its inference
procedure to find the right clustering. We then compare its performance
to that of other stochastic models for graph clustering.

\subsection{Experimental setup}\label{sec3.1}

In order to evaluate the performance of our approach, we applied it on data
generated according to the RSM model. To simplify the parameterization
and facilitate the reproducibility of the experiments, we constrained
the parameters $\bolds\Pi$ and $\bgamma$ to have the following forms:
\[
\bPi= \lleft[\matrix{ \bu& \bv& \cdots& \bv
\cr
\bv& \bu& \ddots& \vdots
\cr
\vdots& \ddots& \ddots& \bv
\cr
\bv& \cdots& \bv& \bu } \rright],\qquad \bgamma= \lleft[\matrix{ \lambda& \varepsilon&
\cdots& \varepsilon
\cr
\varepsilon& \lambda& \ddots& \vdots
\cr
\vdots& \ddots& \ddots& \varepsilon
\cr
\varepsilon& \cdots& \varepsilon& \lambda } \rright],
\]
where $\lambda$, $\varepsilon\in[0,1]$ and $\bu,\bv\in[0,1]^{K}$. With
such a parameterization, the probability $\lambda$ of an edge within a
subgraph is assumed to be common between subgraphs and the probability
$\varepsilon$ of a connection between different subgraphs is also assumed
to be the same for all couples of subgraphs. Similarly, the vector $\bu
$ controls the probability of each edge type between nodes of a same
cluster, whereas $\bv$ defines the edge type probabilities between
nodes of different clusters. We recall that the prior probabilities of
each group within each subgraph are given by the parameter $\balpha
=(\balpha_{1},\ldots,\balpha_{S})$.

Figure~\ref{example} presents an example of a network generated this
way with
parameters $S=2$, $C=3$, $K=3$, $\balpha= \bigl[{\fontsize{8.36pt}{9pt}\selectfont{\matrix{0.1\ \ 0.3\ \ 0.6 \cr
0.6\ \ 0.3\ \ 0.1 }}}\bigr]$, $\lambda=0.6$, $\varepsilon=0.06$, $\bu
=(0.8,0.1,0.1)$ and\vspace*{1pt}
$\bv=(0.1,0.3,0.6)$. This RSM network is made of 30 nodes with $S=2$
subgraphs (indicated by the node forms), $C=3$ types of edges
(indicated by the edge colors) and $K=3$ clusters that have to be
identified in practice (indicated by the node colors).

In order to illustrate, on various situations, that RSM is a
relevant model and that its corresponding inference procedure
provides an accurate estimation of the true clustering structure, we
rely in the following paragraphs on three types of graphs, described
in Table~\ref{graphs}. The three scenarios considered correspond to
different situations ranging from an almost classical setup to a more
specific one. The first scenario considers networks with no
subgraphs ($S=1$) and with a preponderant proportion of edges of
type 1 ($u_{1}=0.8$) and~3 ($u_{3}=0.8$). The second scenario still
considers networks with no subgraphs ($S=1$) but with balanced
proportions of edge types. Finally, the third scenario considers
networks with several subgraphs ($S=3$) and balanced proportions for
edge types. Therefore, the latter case should be the more complex
situation to fit.

%
\begin{table}
\tablewidth=253pt
\caption{Parameter values for the three types of graphs used in the experiments}\label{graphs}
\begin{tabular*}{253pt}{@{\extracolsep{\fill}}lc c  c  c  @{}}
\hline
\textbf{Parameters} & \textbf{Scenario 1} & \textbf{Scenario 2} & \textbf{Scenario 3}\\
\hline
$N$ & 100 & 100 & 100\\
$S$ & \phantom{00}1 & \phantom{00}1 & \phantom{00}3\\
$C$ & \phantom{00}3 & \phantom{00}3 & \phantom{00}3\\
$K$ & \phantom{00}3 & \phantom{00}3 & \phantom{00}3
\\[3pt]
$\balpha$ & (0.3, 0.3, 0.4) & (0.3, 0.3, 0.4) &
$\lleft[\matrix{ 0 & 0.5 & 0.5\cr
0.5 & 0 & 0.5 \cr
0.5 & 0.5 & 0
} \rright]$
\\[11pt]
$\bu$ & (0.8, 0.1, 0.1) & (0.5, 0.45, 0.05) & (0.5, 0.45, 0.05)\\
$\bv$ & (0.1, 0.1, 0.8) & (0.1, 0.45, 0.45) & (0.1, 0.45, 0.45)\\
$\lambda$ & 0.2\phantom{0} & 0.2\phantom{0} & 0.2 \\
$\varepsilon$ & 0.06 & 0.06 & 0.1 \\
\hline
\end{tabular*}
\end{table}
\begin{figure}[t]
\includegraphics{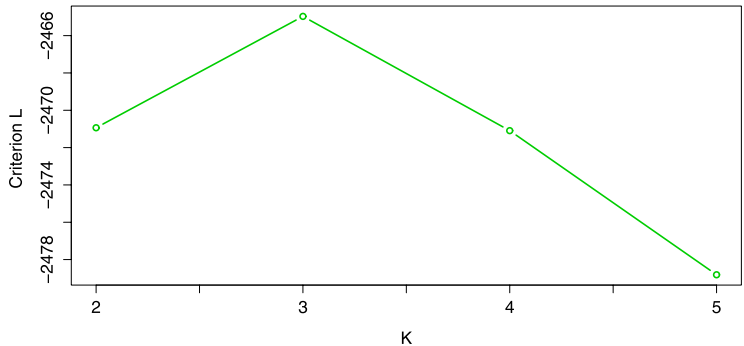}
\caption{Criterion values $\mathcal{L}(q)$ \textit{vs.} number $K$ of
groups for a graph simulated according to scenario~1.}
\label{ARIL}
\end{figure}

The VBEM algorithm with multiple initializations, presented in
Section~\ref{sec2}, is used in the following experiments. For a given value of~$K$, the result with the best
value for $\mathcal{L}(q)$ is chosen among the multiple
initializations. Then, a clustering partition is deduced from the
posterior probabilities $\tau_{ik}$ using the maximum \emph{a posteriori}
(MAP) rule, {that is}, a node is assigned to the group with the
highest posterior probability.

Since our approach aims to search the unobserved clustering partition
of the nodes, we chose here to evaluate the results of our VBEM
algorithm by comparing the resulting partition with the actual one (the
simulated partition). In the clustering community, the adjusted Rand
index (ARI) [\citet{rand1971objective}] serves as a widely accepted
criterion for the difficult task of clustering evaluation. The ARI
looks at all pairs of nodes and checks whether they are classified in
the same group or not in both partitions. As a result, an ARI value
close to 1 means that the partitions are similar and, in our case, that
the VBEM algorithm succeeds to recover the simulated partition.

\subsection{Choice of $K$ and inference results}\label{sec3.2}


In this first simulation study we aim at evaluating the ability of
the lower bound $\mathcal{L}(q)$ to serve as a criterion for
selecting the appropriate number $K$ of clusters. To this end, the
VBEM algorithm for the RSM model was first run on a graph simulated
according to scenario 1 for several values of~$K$. The highest criterion
values among the different initializations obtained for each value of~$K$
are presented in Figure~\ref{ARIL}. The figure indicates that $K=3$
seems to be the appropriate number of groups for the studied network,
which is the actual number of group.

%
\begin{figure}[b]
\includegraphics{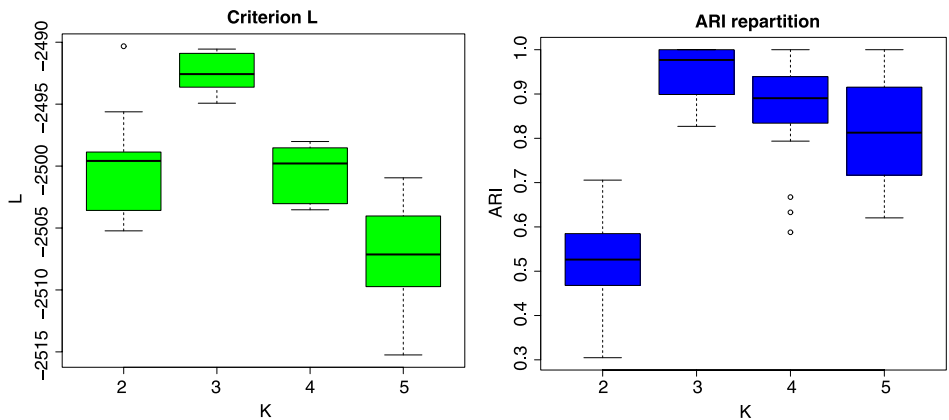}
\caption{Repartition of the criterion (left panel) and ARI (right
panel) over 50 networks
generated with the parameters of the first scenario.}\label{repartition}
\end{figure}

We then replicated this experiment over 50 networks, still simulated
according to scenario 1, for both verifying the consistency of $\mathcal
{L}(q)$ and studying the clustering ability of our approach. Figure~\ref{repartition} shows the repartition of the criterion values (left
panel) as well as the associated ARI values (righ panel). These results
confirm that the lower bound $\mathcal{L}(q)$ is a valid criterion for
selecting the number of groups. One can also observe that the partition
resulting from our VBEM algorithm has, for the selected number of
groups, a good adequation with the actual partition of the data.

\subsection{Comparison with the stochastic block model}\label{sec3.3}

Our second set of experiments compares the performance of RSM to that
of other models on data drawn according to its generative process. We
were interested in the comparison with the following models:
\begin{itemize}
\item\textit{binary SBM} (\textit{presence}): We fit a binary SBM using the R
package \emph{mixer} [\citet{mixer}] on a collapsed version of the data
to conform this specific model. The collapsed data were obtained by
considering only the presence of the edges and not the type of the
edges, {that is}, $\tilde X_{ij} = 0$ if $X_{ij} = 0$ and
$\tilde X_{ij} = 1$ otherwise.

\item\textit{binary SBM} (\textit{type} 1, 2 \textit{or} 3): We fit a binary SBM, still
using the \emph{mixer} package, on the networks defined by taking only
the edges of one type. For instance, the collapsed network for type
$c=1,2,3$ was obtained by\vspace*{1pt} considering only the presence of type $c$
edges, {that is}, $\tilde X_{ij} = 1$ if $X_{ij} = c$ and
$\tilde X_{ij} = 0$ otherwise.

\item\textit{typed SBM}: We consider here a SBM with discrete edges.
Although SBM was originally proposed in \citet{nowicki2001estimation}
with discrete edges, existing softwares only propose to fit a SBM on
binary networks. We therefore had to implement a version of the SBM
which supports typed edges. Note that, in this case, the types of edges
are in $\{0,\ldots,C\}$, where 0 corresponds to the absence of a relation.

\item\emph{RSM}: We run the VBEM algorithm, that we proposed in
Section~\ref{sec2} for the inference of the RSM model, with the available
subgraph partition and with 5 random initializations for each run.
\end{itemize}

%
\begin{table}[b]
\tabcolsep=5pt
\caption{Average ARI values and standard deviations for binary SBM,
typed SBM and RSM according to the three simulation scenarios. The
results are averaged on 25 simulated graphs for each scenario}\label{bestARI}
\begin{tabular*}{\tablewidth}{@{\extracolsep{\fill}}lccc@{}}
\hline
\textbf{Method} & \textbf{Scenario 1} & \textbf{Scenario 2} & \textbf{Scenario 3}\\
\hline
Binary SBM (presence) & 0.001${}\pm{}$0.012 & \phantom{$-$}0.001${}\pm{}$0.013 & \phantom{$-$}0.239${}\pm{}$0.061 \\
Binary SBM (type 1) & 0.976${}\pm{}$0.071 & \phantom{$-$}0.494${}\pm{}$0.233 & $-$0.372${}\pm{}$0.262 \\
Binary SBM (type 2) & 0.001${}\pm{}$0.006 & $-$0.003${}\pm{}$0.006 & \phantom{$-$}0.179${}\pm{}$0.097 \\
Binary SBM (type 3) & 0.959${}\pm{}$0.121 & \phantom{$-$}0.519${}\pm{}$0.219 & \phantom{$-$}0.367${}\pm{}$0.244
\\[3pt]
Typed SBM & 0.694${}\pm{}$0.232 & \phantom{$-$}0.472${}\pm{}$0.339 & \phantom{$-$}0.360${}\pm{}$0.162
\\[3pt]
RSM & \textbf{1.000${}\bolds{\pm}{}$0.000} & \phantom{$-$}\textbf{0.981${}\bolds{\pm}{}$0.056} & \phantom{$-$}\textbf{0.939${}\bolds{\pm}{}$0.097} \\
\hline
\end{tabular*}
\end{table}

Table~\ref{bestARI} presents the average ARI values and standard
deviations on 50 simulated graphs for each scenario and with binary SBM,
typed SBM and RSM. We point out that the inference is done with the
actual number of clusters and this for each method. One can observe
that, for the first scenario, the
binary SBM based on the link presences and the type 2 SBM always fails,
whereas type 1, type 3 and typed SBM work pretty well. Those behaviors
can be explained by the nature of scenario~1, which is a rather easy
situation with no subgraphs and a predominant presence of type 1 and
type 2 links. However, we can remark that it seems easier in this case
to fit a binary SBM on type 1 or type 2 edges than to fit a typed
SBM. This is due to the high discriminative power of type 1 and type 2
edges in this specific scenario. Let us also remark that RSM works
perfectly here even though the network does not contain any subgraphs.

Regarding scenario 2, which considers a situation where there is still
no subgraphs but with more balanced proportions of the different edge
types, one can first notice that binary SBM and type 2 SBM fail once
again. The type 1 and type 3 SBM have now a behavior closer to the
one of typed SBM, whereas RSM gives very accurate results once
again. Finally, scenario 3 considers a RSM-type network, {that is},
with several subgraphs, and all SBM-based algorithms are significantly
outperformed by RSM, which succeeds in exploiting both the information
carried by different edge types and by the different subgraphs. To
summarize, the RSM model and its associated VBEM algorithm turn out to
be effective on situations ranging from classical setups without
subgraphs to complex scenarios with subgraphs and typed edges.

\section{Ecclesiastical network}\label{sec4} \label{sececclNet}

This section now focuses on applying the RSM model to the
ecclesiastical network, that we briefly
described in the \hyperref[sec1]{Introduction} and that initially motivated this
work, and on analyzing its results from the historical point of
view.

Please note that the ecclesiastical network along with a
file giving the kingdoms of all vertices in the network and an R code
implementing the variational inference approach for the RSM model
are provided as supplementary materials in \citet{suppsuppA}.

\subsection{Description of the data}\label{sec4.1}

The relational data considered in this section were mainly built from
written acts of ecclesiastical councils
that took place in Merovingian Gaul during the 6th century. A council
is an ecclesiastical meeting, usually called by a bishop, where issues
regarding the Church or the faith are addressed. However, since 511,
kings could also call for a council to discuss some political,
judiciary or legal issues, and that laics (kings, dukes or counts, e.g.) would attend. During the 6th century, 46 councils took place
in Gaul. Although there were mostly local or regional councils,
attended by individuals from a specific ecclesiastical province, there
were some national councils convened under the authority of a king.

The composition of these councils is known thanks to the acts written
at the end of the meeting, and which were signed by all attending
members. In addition to the council acts, we used narrative texts
(among which is the famous \emph{Ten History Books} by Gregory of
Tours), hagiographies or letters which also describe these councils.
The network, that took over 18 months to build from these historical
sources, contains $N=1331$ individuals who held one or several offices
in Gaul between the years 480 and 614, and who we know to have been
related or to have met during their lifetime.

The council acts and the other historical sources allowed also to
qualify the type of the relationship between the individuals involved
in the network. However, the scarcity of the sources only allowed for
an approximate characterization of these relationships. As a
consequence, $C=4$ relation types were qualified and the relationships
can be either positive, negative, variable or neutral (when the type
was unknown). For instance, a positive relationship may describe an
agreement between two clergymen on a question of faith, whereas a
negative one may be a disagreement on such a question. Variable
relationships usually correspond to relationships which change over time.

Using the different sources, it was also possible to obtain additional
information on the individuals. In particular, the geographical
positions of the offices hold by the clergymen or the laics allowed us
to split the network into $S=6$ subgraphs. Those 6 subgraphs correspond
to the geographical partition of the Gaul at this period (the kingdoms
of Neustria, Austrasia and Burgondy, and the provinces of Aquitaine and
Provence), completed with an additional subgraph for individuals for
whom the information was not available. We also recorded the social
positions of the individuals in order to be able to interpret afterward
the clusters found by our method. These social positions can be, for
instance, ecclesiastical positions (bishops, deacon, archdeacon, abbot,
priest$,\ldots$) or titles of nobility (king, queen, duke, earl$,\ldots$).

%
\begin{figure}[b]

\includegraphics{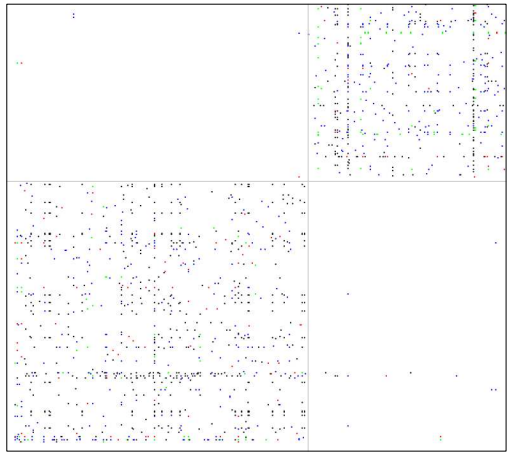}

\caption{Adjacency matrix for the kingdom of Neustria (left block) and
the province of Provence (right block). The dot colors indicate the
type or relationships: red${}={}$``negative,'' green${}={}$``variable,'' black${}={}$``neutral''
and blue${}={}$``positive.'' Zoom on the paper electronic version for details.}
\label{AdjMatrix12}
\end{figure}

%
\begin{figure}

\includegraphics{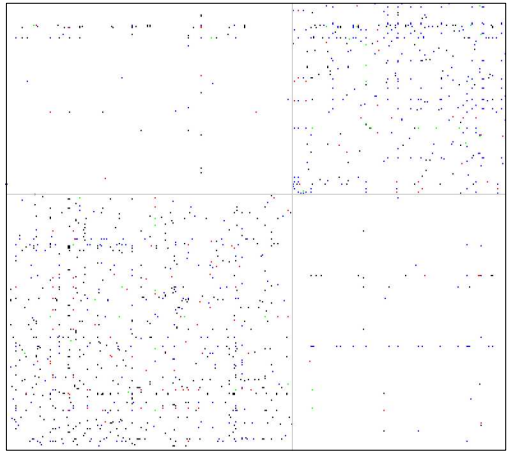}

\caption{Adjacency matrix for the kingdoms of Austrasia (left block)
and Burgundy (right block). The dot colors indicate the type or
relationships: red${}={}$``negative,'' green${}={}$``variable,'' black${}={}$``neutral''
and blue${}={}$``positive.'' Zoom on the paper electronic version for details.}
\label{AdjMatrix45}
\end{figure}

To summarize, the network is made of $N=1331$ individuals split into
$S=6$ subgraphs and whose relationships can be of $C=4$ difference
types. Figures \ref{AdjMatrix12}~and~\ref{AdjMatrix45} show some parts
of the whole adjacency matrix associated to the network where the dot
colors indicate the type or relationships. The whole adjacency matrix
is provided in a zoomable pdf file as supplementary material [\citet{suppsuppA}]. We expect
the statistical analysis with RSM of this network to help us understand
how the behavior of an individual can be modeled through their
belonging to a group. The use of a probabilistic approach, instead of a
deterministic one, makes particular sense here since at least a part of
the historical sources are subject to caution due to their nature and
age. In History, this kind of approach is more common to modernists or
contemporarists than to medievists who rarely have access to this kind
of data. Let us finally notice that a ``source effect'' is expected due
to the possible overrepresentation in our sources of some places
(Neustria by Gregory of Tours or Austrasia by Fredegar) or some
individuals (in letters or hagiographies).\looseness=1

\subsection{Results}\label{sec4.2}

The VBEM algorithm that we proposed to infer the RSM model was run on
the network defined by these relations, where the subgraphs
are the provinces in which the individuals lived (Aquitaine, Austrasia,
Burgondy, Neustria, Provence or Unknown). The use of the lower bound
$\mathcal{L}(q)$ allowed us to find 6 clusters.

%
\begin{figure}[t]
\includegraphics{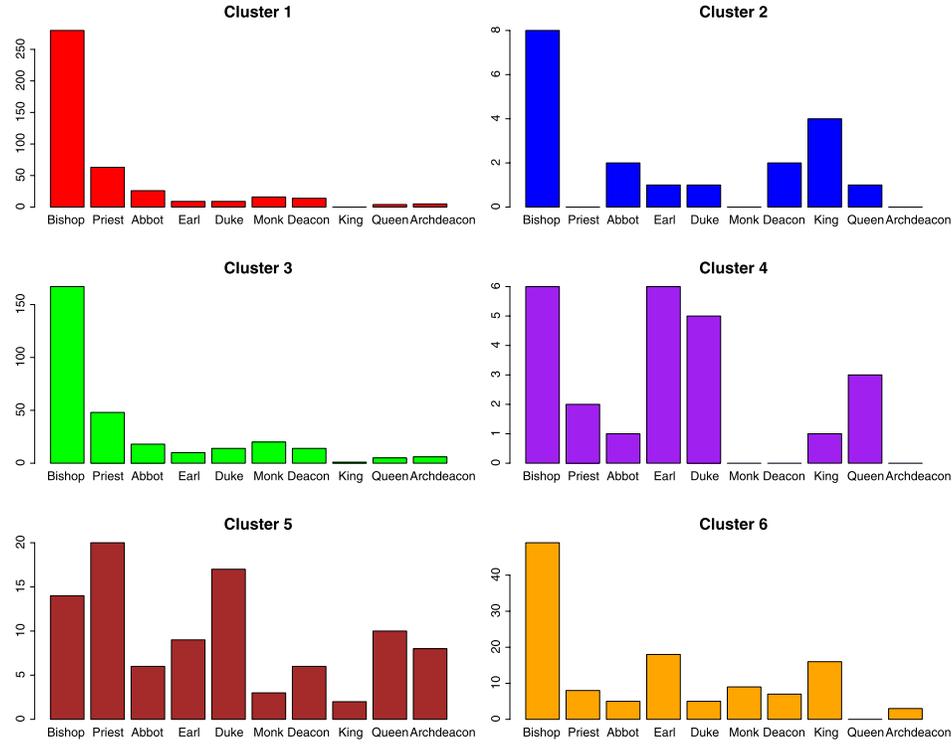}
\caption{Repartition of the different social positions in the found
clusters (restricted to the 10 most frequent positions).}\label{Clusters}
\end{figure}

To give some insight into the nature of the found clusters, Figure~\ref{Clusters} presents the repartition of the different social positions
in the clusters. In view of these results, some historical comments can
be done. First, clusters 1 and 3 appear to be made of the people who
would attend local assemblies, provincial or diocesan councils. The
council of Arles, which took place in 554, would have had the same kind
of composition as cluster 3, while that of Auxerre, in 585, could well
represent cluster~1. Second, clusters 4 and 5 are more characteristic
of aristocratic assemblies, such as the council of Orange in 529.
Third, clusters 2 and 6 have the same compositions as councils
concerned with more political issues (those usually convened by a
king). Such a council took place in Orleans in 511. Let us, however,
notice that cluster 2 is composed of very few individuals, which might
hurt the relevance of its interpretation. Also, we might be able to
further our understanding of the composition of these clusters by
taking into account the similarity of certain social positions (such as
``duke'' and ``earl'').

%
\begin{figure}[t]
\includegraphics{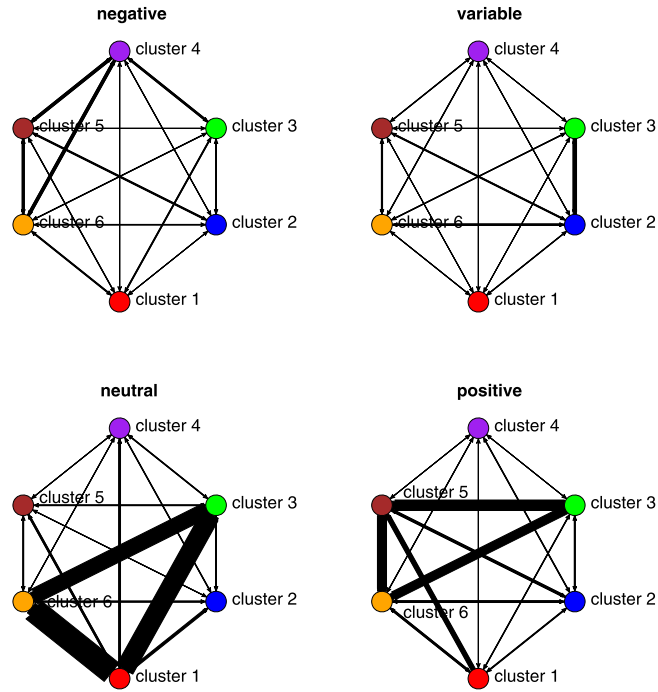}
\caption{Relations between the 6 found clusters (parameter $\bPi$) for
each relation type: negative, variable, neutral and positive. For
visualization purposes, the relation weights have been normalized
according to relation types.}
\label{relations}
\end{figure}

The relations between the different clusters, described by the
parameter $\bPi$ and shown in Figure~\ref{relations}, inform us
further. Although the limitations expressed above about the roughness
of the relation types still apply, they nevertheless provide us with
interesting elements to confirm the coherence of the proposed model.
First, it is natural that we should find ``neutral'' relations at the
local level, between clusters 1, 3 and 6. Indeed, local assemblies were
the less documented ones in our sources. On the other hand, the links
between high level individuals are better known, because councils used
to settle conflicts between aristocrats, which explains the presence of
``negative'' and ``variable'' relations. Finally, the positive
relations between cluster 3, 5 and 6 could represent the personal
friendships documented by a collection of letters between bishops.

%
\begin{figure}
\includegraphics{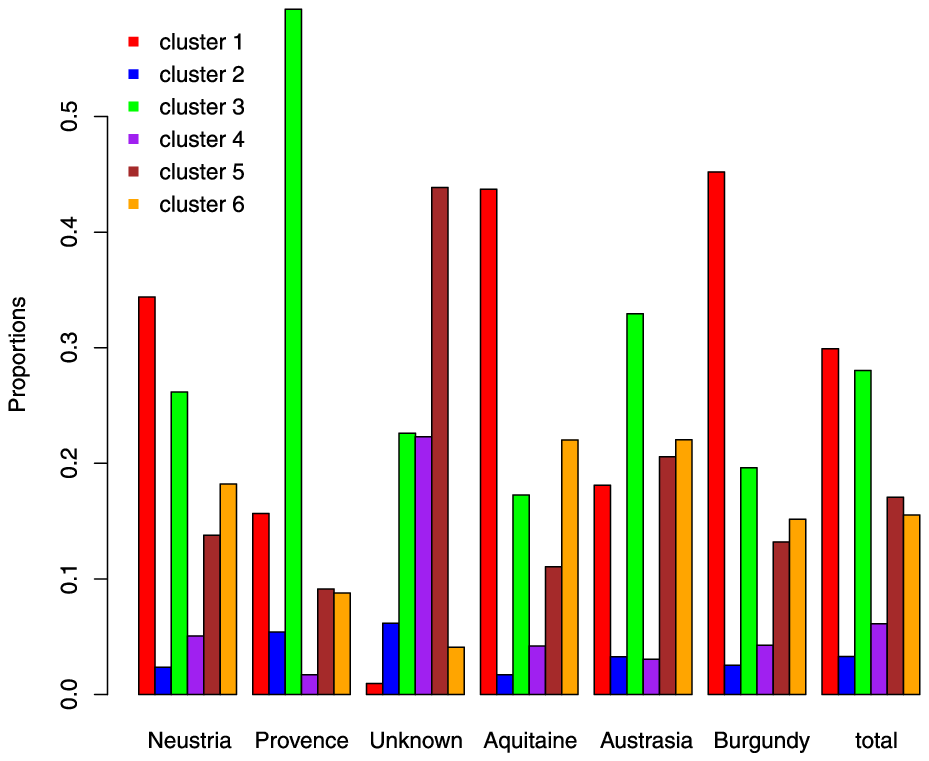}
\caption{Repartition of the 6 found clusters in the
different Provinces (parameter $\balpha$).}
\label{Provinces}
\end{figure}

%
\begin{figure}[b]
\includegraphics{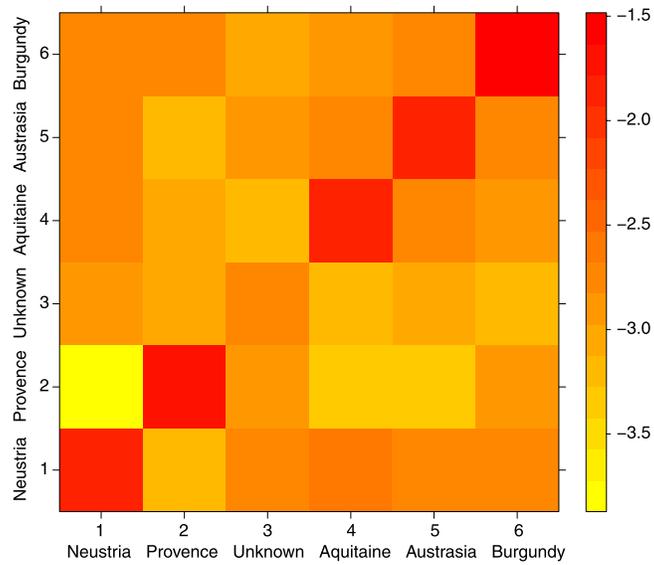}
\caption{Estimated values for the parameter $\bgamma$ (in log scale).}\label{comm}
\end{figure}

After having described the political background represented by each of
the clusters, we can compare the organization of the different regions.
Figure~\ref{Provinces} presents the cluster repartition (parameter
$\balpha$) in the different provinces. One can observe that the clergy
and noblemen of the different regions were concerned with very
different issues: Provence and Burgundy were more concerned with local
questions (clusters 1 and 3), and less with political ones (clusters 2
and 6). The clusters concerned both with local (clusters 1 and 3) and
high level (cluster 6) questions are represented in Aquitaine.
Conversely, all levels of power are represented in Neustria. This could
be the result of a ``source effect,'' as mentioned above. Let us also
notice that the council structures seem similar in Austrasia and
Aquitaine: sovereigns (kings and queens) are involved in the Church and
frequently convene councils in order to discuss political questions.

Some of these observations are confirmed by the estimate of parameter
$\bgamma$, which is given in a log scale by Figure~\ref{comm}. First,
it shows a greater frequency of relations between Aquitaine and
Neustria, which comes both from a geographical and political proximity
(Aquitaine is absorbed into Clovis' kingdom in 507, then divided and
absorbed by Neustria in 511). One can also see there another example of
``source'' effect, as our main source, Gregory of Tours, was bishop in
Neustria and raised in Aquitaine (next to his uncle, the bishop of
Clermont), which gave him a good knowledge of both provinces. More
enlightening is the relative disconnection of Burgondy and Provence,
especially in regard to the provinces of Austrasia and Neustria, both
heavily connected.\looseness=1

\subsection{Conclusion from the historical point of view}\label{sec4.3}

A first analysis of the results of the RSM model confirms two
well-known general facts. Indeed, our results confirm the preponderance
of local assemblies in 6th century Gaul and the ``source effect.''
Nevertheless, further analysis of the found clusters and their
relations yields a better understanding of the period. In particular,
the composition of the found clusters reflect different archetypes of
councils and different levels of political concerns. Our results have
also highlighted that the types of concerns of each province are
closely related to the frequency of their communications with others.

Two limitations to these results remain, however. First, we are limited
by the scarcity of the historical documentation. It would be
interesting to see whether the use of more precise types of relations
(ecclesiastical or secular, through which media$,\ldots$) could improve
the results. Second, it would also be interesting for the model to
take into account temporal evolutions of the relations and
clusters. Indeed, one aspect of the data which is currently not
addressed by the results of RSM is its temporality. Nevertheless, this
lack seems to have a limited impact here since all clusters exhibit the
same distribution of individuals over time, reflecting the higher
concentration of information in years 550 to 600 (when numerous
conflicts were settled by councils: Paris 577, Chalon 579, Berny 580,
Lyon 581$,\ldots$). The repartition of the different kinds of powers then
seems to change little over time on this short period.

\section{Conclusion and further work}\label{sec5}

In this work we proposed a new stochastic graph model, the random
subgraph model, to deal with networks where a vertex behavior is
influenced by an observed partition variable. We derived a variational
Bayes EM algorithm to infer the model parameters from data and applied
it to an ecclesiastical network from Merovingian Gaul. The results of
the fitted RSM enlightened us on the different levels of power present
at this time in Gaul, and on the different power structures of
different regions. Let us highlight that the RSM
model allows in addition the comparison of subgraphs through the model
parameters, in particular, the cluster proportions. We also would like
to mention that networks with typed edges and subgraphs can be
encountered in many application fields (such as biology, economics,
archeology$,\ldots$) and the RSM model should be useful in these contexts
as well.

One aspect, however, that RSM does not currently address is the
temporality of the data. Since this aspect can be found in many of
the data sets we wish to apply the RSM model to, we believe that a
natural continuation of this work would be a dynamic extension of the
RSM model. Moreover, we plan to introduce a~Chinese restaurant process
on the latent cluster structure in order to automatically estimate the
number of clusters while clustering the vertices. Finally, we would
like to consider the problem of visualizing such networks with typed
edges and known subgraphs.


\begin{appendix}
\section*{Appendix: Variational Bayes}\label{app}
In this final section we detail the computations that lead to the
update rules
given in Section~\ref{sec2}, and provide an explicit expression of the criterion
$\mathcal{L}(q)$.

%
\begin{prop}
The complete data log-likelihood the RSM model is given~by
\begin{eqnarray*}
&& \log p(\X,\A,\Z,\balpha,\bgamma,\bPi)
\\
&&\qquad = \sum_{i\neq j}^{N}
\sum_{c=1}^{C}\sum
_{k,l}^{K} \bigl\{\delta(X_{ij}=c)
Z_{ik}Z_{jl}\log(\Pi _{klc}) \bigr\}
\\
&&\quad\qquad{}+ \sum_{i=1}^{N}\sum
_{k=1}^{K}Z_{ik}\log(\alpha_{r_i,k})
\\
&&\quad\qquad{}+\sum_{i\neq j}^{N} \bigl\{A_{ij}
\log(\gamma_{r_i,r_j})+(1-A_{ij})\log (1-\gamma_{r_i,r_j})
\bigr\}
\\
&&\quad\qquad{}+\sum_{s=1}^{S}\log p(
\balpha_{s}) + \sum_{r,s}^{S}\log
p(\gamma _{rs}) + \sum_{k,l}^{K}
\log p(\bPi_{kl}).
\end{eqnarray*}
\end{prop}

\begin{pf}
\begin{eqnarray*}
&&\log p(\X,\A,\Z,\balpha,\bgamma,\bPi)
\\
&&\qquad = \log p(\X|\A,\Z,\bPi) + \log p(\A|\bgamma) + \log p(\Z|\balpha) + \log p(\balpha) + \log p(
\bgamma)
\\
&&\quad\qquad{}+ \log p(\bPi)
\\
&&\qquad = \sum_{i\neq j}^{N} \sum
_{c=1}^{C}\sum_{k,l}^{K}
\bigl\{\delta (X_{ij}=c) Z_{ik}Z_{jl}\log(
\Pi_{klc}) \bigr\}
\\
&&\quad\qquad{}+ \sum_{i=1}^{N}\sum
_{k=1}^{K}Z_{ik}\log(\alpha_{r_i,k})
\\
&&\quad\qquad{}+\sum_{i\neq j}^{N} \bigl\{A_{ij}
\log(\gamma_{r_i,r_j})+(1-A_{ij})\log (1-\gamma_{r_i,r_j})
\bigr\}
\\
&&\quad\qquad{}+\sum_{s=1}^{S}\log p(
\balpha_{s}) + \sum_{r,s}^{S}\log
p(\gamma_{rs}) + \sum_{k,l}^{K}
\log p(\bPi_{kl}).
\end{eqnarray*}\upqed
\end{pf}

%
\begin{prop}
The VBEM update step for the distribution $q(\gamma_{rs})$ is given~by
\[
q(\gamma_{rs})=\operatorname{Beta}(\gamma_{rs};a_{rs},b_{rs})\qquad\forall(r,s)\in \{1,\ldots,S\}^{2},
\]
where
\[
a_{rs}=a_{rs}^{0}+\sum
_{r_i=r,r_j=s}(A_{ij})
\]
and
\[
b_{rs}=b_{rs}^{0}+\sum
_{r_i=r,r_j=s}(1-A_{ij}).
\]
\end{prop}

\begin{pf}
\begin{eqnarray*}
&& \log q(\gamma_{rs})
\\[-1pt]
&&\qquad =\mathbb{E}_{\Z,\balpha,\bgamma^{\setminus
rs},\bPi}\bigl[\log p(\X,\A,\Z,
\balpha,\bgamma,\bPi)\bigr]+\kappa
\\[-1pt]
&&\qquad =\sum_{r_i=r,r_j=s} \bigl\{A_{ij}\log(
\gamma_{rs})+(1-A_{ij})\log (1-\gamma_{rs}) \bigr\},
\\[-1pt]
&& \log p(\gamma_{rs}) + \kappa
\\[-1pt]
&&\qquad =\sum_{r_i=r,r_j=s} \bigl\{A_{ij}\log(
\gamma_{rs})+(1-A_{ij})\log (1-\gamma_{rs}) \bigr\},
\\[-1pt]
&&\bigl(a_{rs}^{0}-1\bigr)\log(\gamma_{rs})+
\bigl(b_{rs}^{0}-1\bigr)\log(1-\gamma _{rs})+\kappa
\\[-1pt]
&&\qquad =\biggl(a_{rs}^{0}-1+\sum_{r_i=r,r_j=s}A_{ij}
\biggr)\log(\gamma_{rs})
\\[-1pt]
&&\quad\qquad{}+\biggl(b_{rs}^{0}-1+\sum_{r_i=r,r_j=s}(1-A_{ij})
\biggr)\log(1-\gamma_{rs})+\kappa,
\end{eqnarray*}
where $\kappa$ is a constant term. Hence, the functional form of the
variational approximation $q(\gamma_{rs})$ corresponds to a Beta
distribution with updated hyperparameters:
\[
a_{rs}=a_{rs}^{0}+\sum
_{r_i=r,r_j=s}(A_{ij})
\]
and
\[
b_{rs}=b_{rs}^{0}+\sum
_{r_i=r,r_j=s}(1-A_{ij}).
\]\upqed
\end{pf}

%
\begin{prop}
The VBEM update step for the distribution $q(\Z_{i})$ is given by
\[
q(\Z_{i})=\mathcal{M}(\Z_{i};1,\btau_{i})\qquad \forall i,
\]
where
\begin{eqnarray*}
\tau_{ik} &\propto& \exp \Biggl(\psi(\chi_{r_i,k})-\psi\Biggl(
\sum_{l=1}^{K}\chi_{r_i,l}\Biggr)
\Biggr)
\\[-1pt]
&&{}+\exp \Biggl\{\sum_{j\ne i}^{N} \sum
_{c=1}^{C}\sum_{l=1}^{K}
\delta(X_{ij}=c)\tau_{jl} \Biggl(\psi(\Xi_{klc})-\psi
\Biggl(\sum_{u=1}^{C}\Xi_{klu}
\Biggr) \Biggr) \Biggr\}
\\[-1pt]
&&{}+\exp \Biggl\{\sum_{j\ne i}^{N} \sum
_{c=1}^{C}\sum_{l=1}^{K}
\delta(X_{ji}=c)\tau_{jl} \Biggl(\psi(\Xi_{lkc})-\psi
\Biggl(\sum_{u=1}^{C}\Xi_{lku}
\Biggr) \Biggr) \Biggr\}.
\end{eqnarray*}
\end{prop}

\begin{pf}
\begin{eqnarray*}
\log q(\Z_i)&=&\mathbb{E}_{\Z^{\setminus i},\balpha,\bgamma,\bPi} \bigl[\log p(\X,\A,\Z,
\balpha,\bgamma,\bPi)\bigr]+\kappa
\\[-2pt]
&=&\mathbb{E}_{\Z^{\setminus i},\bPi} \Biggl[\sum_{j=1}^{N}
\sum_{c=1}^{C} \Biggl\{\delta(X_{ij}=c)
\sum_{k,l}^{K}Z_{ik}Z_{jl}
\log(\Pi _{klc}) \Biggr\} \Biggr]
\\[-2pt]
&&{}+\mathbb{E}_{\Z^{\setminus i},\bPi} \Biggl[\sum_{j=1}^{N}
\sum_{c=1}^{C} \Biggl\{\delta(X_{ji}=c)
\sum_{k,l}^{K} Z_{jk}Z_{il}
\log(\Pi_{klc}) \Biggr\} \Biggr]
\\[-2pt]
&&{}+ \mathbb{E}_{\Z^{\setminus i},\balpha} \Biggl[ \sum_{k=1}^{K}
Z_{ik}\log(\alpha_{r_i,k}) \Biggr]+\kappa
\\[-2pt]
&=&\sum_{k=1}^{K}Z_{ik}
\mathbb{E}_{\balpha} \bigl[\log(\alpha_{r_i,k})\bigr]
\\[-2pt]
&&{}+\sum_{j=1}^{N}\sum
_{c=1}^{C}\sum_{l,k}^{K}Z_{ik}
\delta (X_{ij}=c)\mathbb{E}_{Z^{\setminus
i},\bPi} \bigl[Z_{jl}\log(
\Pi_{klc})\bigr]
\\[-2pt]
&&{}+\sum_{j=1}^{N}\sum
_{c=1}^{C}\sum_{l,k}^{K}Z_{ik}
\delta (X_{ji}=c)\mathbb{E}_{Z^{\setminus
i},\bPi} \bigl[Z_{jl}\log(
\Pi_{lkc})\bigr]+\kappa
\\[-2pt]
&=&\sum_{k=1}^{K}Z_{ik} \Biggl(
\psi(\chi_{r_i,k})-\psi\Biggl(\sum_{l=1}^{K}
\chi _{r_i,l}\Biggr) \Biggr)
\\[-2pt]
&&{}+\sum_{k=1}^{K}Z_{ik} \Biggl\{
\sum_{j\ne i}^{N} \sum
_{c=1}^{C}\sum_{l=1}^{K}
\delta(X_{ij}=c)\tau_{jl} \Biggl(\psi(\Xi_{klc})-\psi
\Biggl(\sum_{u=1}^{C}\Xi_{klu}
\Biggr) \Biggr) \Biggr\}
\\[-2pt]
&&{}+ \sum_{k=1}^{K}Z_{ik} \Biggl\{
\sum_{j\ne i}^{N} \sum
_{c=1}^{C}\sum_{l=1}^{K}
\delta(X_{ji}=c)\tau_{jl} \Biggl(\psi(\Xi_{lkc})-\psi
\Biggl(\sum_{u=1}^{C}\Xi_{lku}
\Biggr) \Biggr) \Biggr\}
\\[-3pt]
&&{} +\kappa,
\end{eqnarray*}
where $\kappa$ is a constant term. Hence, the functional form of the
variational approximation $q(\Z_i)$ corresponds to a multinomial
distribution, with updated parameters:
\begin{eqnarray*}
\tau_{ik} &\propto& \exp \Biggl(\psi(\chi_{r_i,k})-\psi\Biggl(
\sum_{l=1}^{K}\chi_{r_i,l}\Biggr)
\Biggr)
\\[-2pt]
&&{}+\exp \Biggl\{\sum_{j\ne i}^{N} \sum
_{c=1}^{C}\sum_{l=1}^{K}
\delta(X_{ij}=c)\tau_{jl} \Biggl(\psi(\Xi_{klc})-\psi
\Biggl(\sum_{u=1}^{C}\Xi_{klu}
\Biggr) \Biggr) \Biggr\}
\\[-2pt]
&&{}+\exp \Biggl\{\sum_{j\ne i}^{N} \sum
_{c=1}^{C}\sum_{l=1}^{K}
\delta(X_{ji}=c)\tau_{jl} \Biggl(\psi(\Xi_{lkc})-\psi
\Biggl(\sum_{u=1}^{C}\Xi_{lku}
\Biggr) \Biggr) \Biggr\}.
\end{eqnarray*}\upqed
\end{pf}

%
\begin{prop}
The VBEM update step for the distribution $q(\balpha_{s})$ is given
by
\[
q(\balpha_{s})=\operatorname{Dir}(\balpha_{s};
\chi_{s})\qquad \forall s\in\{1,\ldots,S\},
\]
where
\[
\chi_{sk}=\chi_{sk}^0+ \sum
_{i=1}^{N}\delta(r_{i}=s)
\tau_{ik}\qquad\forall k\in\{1,\ldots,K\}.
\]
\end{prop}

\begin{pf}
\begin{eqnarray*}
\log q(\balpha_s)&=&\mathbb{E}_{\Z,\balpha^{\setminus s},\bgamma,\bPi} \bigl[\log p(\X,\A,\Z,
\balpha,\bgamma,\bPi)\bigr]+\kappa
\\
&=&\mathbb{E}_{\Z,\balpha^{\setminus s}} \Biggl[ \sum_{i=1}^{N}
\sum_{k=1}^{K} Z_{ik}\log(
\alpha_{r_i,k})\Biggr] +\log p(\balpha_{s})+\kappa
\\
&=&\sum_{k=1}^{K}\sum
_{i=1}^{N}\delta(r_i=s)\log(
\alpha_{sk})\mathbb {E}_{\Z} [Z_{ik}] + \sum
_{k=1}^K\log(\alpha_{sk}) \bigl(
\chi_{sk}^{0}-1\bigr)+\kappa
\\
&=&\sum_{k=1}^{K}\log(\alpha_{sk})
\Biggl\{\chi_{sk}^{0}-1+ \sum_{i=1}^{n}
\delta(r_i=s)\tau_{ik} \Biggr\}+\kappa,
\end{eqnarray*}
where $\kappa$ is a constant term. Hence, the functional form of the
variational approximation $q(\balpha_s)$ corresponds to a Dirichlet
distribution with updated hyperparameters:
\[
\chi_{sk}=\chi_{sk}^0+ \sum
_{i=1}^{N}\delta(r_{i}=s)
\tau_{ik}\qquad\forall k\in\{1,\ldots,K\}.
\]\upqed
\end{pf}

%
\begin{prop}
The VBEM update step for the distribution $q(\bPi_{kl})$ is given by
\[
q(\bPi_{kl})=\operatorname{Dir}(\bPi_{kl};\Xi_{kl})\qquad
\forall(k,l)\in\{1,\ldots,K\}^{2},
\]
where
\[
\Xi_{klc}=\Xi_{klc}^{0}+\sum
_{i\neq j}^{N}\delta(X_{ij}=c)
\tau_{ik}\tau _{jl}\qquad\forall c\in\{1,\ldots,C\}.
\]
\end{prop}
\begin{pf}
\begin{eqnarray*}
\log q(\bPi_{k,l})&=&\mathbb{E}_{\Z,\balpha,\bgamma,\bPi^{\setminus
kl}} \bigl[\log p(\X,\A,\Z,
\balpha,\bgamma,\bPi)\bigr]+\kappa
\\
&=& \mathbb{E}_{\Z,\bPi^{\setminus kl}} \Biggl[\sum_{i\neq j}^{N}
\sum_{c=1}^{C} \delta(X_{ij}=c)Z_{ik}Z_{jl}
\log(\Pi_{klc})\Biggr] + \log p(\bPi_{kl})+\kappa
\\
&=&\sum_{c=1}^C\log(\Pi_{klc})
\Biggl\{\sum_{i\neq j}^{N}\delta
(X_{ij}=c)\tau_{ik}\tau_{jl} \Biggr\}
\\
&&{} +\sum
_{c=1}^C\log(\bPi_{klc}) \bigl(
\Xi_{klc}^{0}-1\bigr)+\kappa
\\
&=& \sum_{c=1}^C\log(\Pi_{klc})
\Biggl\{\Xi_{klc}^{0}-1 +\sum_{i\neq j}^{N}
\delta(X_{ij}=c)\tau_{ik}\tau_{jl} \Biggr\} +
\kappa,
\end{eqnarray*}
where $\kappa$ is a constant term. Hence, the functional form of the
variational approximation $q(\bPi_{kl})$ corresponds to a
Dirichlet distribution with updated hyperparameters:
\[
\Xi_{klc}=\Xi_{klc}^{0}+\sum
_{i\neq j}^{N}\delta(X_{ij}=c)
\tau_{ik}\tau _{jl}\qquad\forall c\in\{1,\ldots,C\}.
\]\upqed
\end{pf}

%
\begin{prop}
When computed right after the M step, the lower bound of the
marginal log-likelihood is given by
\begin{eqnarray*}
\mathcal{L}(q)&=&\sum_{r,s}^{S}\log\biggl(
\frac
{B(a_{rs},b_{rs})}{B(a_{rs}^{0},b_{rs}^{0})}\biggr) +\sum_{s=1}^{S}\log
\biggl(\frac{C(\bolds \chi_{s})}{C(\bolds
\chi_{s}^{0})}\biggr)
\\
&&{} + \sum_{k,l}^{K}
\log\biggl(\frac{C(\bolds
\Xi_{kl})}{C(\bolds
\Xi_{kl}^{0})}\biggr)-\sum_{i=1}^{N}
\sum_{k=1}^{K}\tau_{ik}\log(
\tau_{ik}),
\end{eqnarray*}
where
$C(x)=\frac{\prod_{d=1}^{D}\Gamma(x_{d})}{\Gamma(\sum_{d=1}^{D}x_{d})}$
if $x\in\mathbb{R}^{D}$ and
$B(a,b)=\frac{\Gamma(a)\Gamma(b)}{\Gamma(a+b)},\forall(a,b)\in\mathbb{R}^{2}$.
\end{prop}

\begin{pf}
The lower bound is given by
\[
\mathcal{L}(q)=\mathbb{E}_{\Z,\balpha,\bgamma,\bPi}\biggl[\log\biggl(\frac{p(\X,\A,\Z,\balpha,\bgamma,\bPi)}{q(\Z,\balpha,\bgamma,\bPi)}\biggr)\biggr],
\]
where
\begin{eqnarray*}
\log\biggl(\frac{p(\X,\A,\mathbf Z,\bolds\alpha,\bolds
\gamma,\bolds\Pi)}{q(\mathbf Z,\bolds\alpha,\bolds\gamma,\bolds
\Pi)}\biggr) &=&\sum_{i \neq j}^{N}
\bigl\{ A_{ij}\log(\gamma_{r_i,r_j})+(1-A_{ij})\log(1-
\gamma_{r_i,r_j}) \bigr\}
\\
&&{}+\sum_{i \neq j}^{N} \sum
_{c=1}^{C}\sum_{k,l}^{K}
\bigl\{ \delta(X_{ij}=c) Z_{ik}Z_{jl}\log(
\Pi_{klc}) \bigr\}
\\
&&{}+\log\biggl(\frac{p(\Z,\balpha,\bgamma,\bPi)}{q(\Z,\balpha,\bgamma,\bPi)}\biggr)\vadjust{\goodbreak}
\end{eqnarray*}
and
\begin{eqnarray*}
&& \log\biggl(\frac{p(\Z,\balpha,\bgamma,\bPi)}{q(\Z,\balpha,\bgamma,\bPi
)}\biggr)
\\
&&\qquad =\log\biggl(\frac{p(\Z|\balpha)}{q(\Z)}\biggr)+\log
\biggl(\frac{p(\balpha,\bgamma,\bPi
)}{q(\balpha,\bgamma,\bPi)}\biggr)
\\
&&\qquad =\sum_{i=1}^{N}\sum
_{k=1}^{K}Z_{ik}\log\biggl(
\frac{\alpha_{r_i,k}}{\tau_{ik}}\biggr) + \sum_{r,s}^{S}
\log\biggl(\frac{\operatorname{Beta}(\gamma
_{rs};a_{rs}^{0},b_{rs}^{0})}{\operatorname{Beta}(\gamma_{rs};a_{rs},b_{rs})}\biggr)
\\
&&\quad\qquad{}+ \sum_{s=1}^{S}\log\biggl(
\frac{\operatorname{Dir}(\bolds \alpha_s;\bolds
\chi_{s}^0)}{\operatorname{Dir}(\bolds \alpha_s;\bolds \chi_{s})}\biggr) + \sum_{k,l}^{K}
\log\biggl(\frac{\operatorname{Dir}(\bolds \Pi_{kl};\bolds
\Xi_{kl}^{0})}{\operatorname{Dir}(\bolds\Pi_{kl};\bolds\Xi_{kl})}\biggr)
\\
&&\qquad =\sum_{i=1}^{N}\sum
_{k=1}^{K}Z_{ik}\log\biggl(
\frac{\alpha_{r_i,k}}{\tau_{ik}}\biggr)+ \sum_{r,s}^{S}
\log\biggl(\frac
{B(a_{rs},b_{rs})}{B(a_{rs}^{0},b_{rs}^{0})\gamma
_{rs}^{a_{rs}-a_{rs}^{0}}(1-\gamma_{rs})^{b_{rs}-b_{rs}^{0}}}\biggr)
\\
&&\quad\qquad{}+\sum_{s=1}^{S}\log\biggl(
\frac{C(\bolds \chi_{s})}{C(\bolds
\chi_{s}^{0})\prod_{k=1}^{K}\alpha_{k}^{\chi_{sk} -
\chi_{sk}^{0}}}\biggr) + \sum_{k,l}^{K}
\log\biggl(\frac{C(\bolds\Xi
_{kl})}{C(\bolds\Xi_{kl}^{0})\prod_{c=1}^{C}\Pi_{klc}^{\Xi_{klc}-\Xi
_{klc}^{0}}}\biggr).
\end{eqnarray*}
If $x\in\mathbb{R}^{D}$, then
$C(x)=\frac{\prod_{d=1}^{D}\Gamma(x_{d})}{\Gamma(\sum_{d=1}^{D}x_{d})}$,
where $\Gamma(\cdot)$ is the gamma function. Moreover, if
$(a,b)\in\mathbb{R}^{2}$, then $B(a,b)=\frac{\Gamma(a)\Gamma(b)}{\Gamma(a+b)}$.
Finally,
\begin{eqnarray*}
&& \log\biggl(\frac{p(\X,\A,\Z,\bolds\alpha,\bolds
\gamma,\bolds\Pi)}{q(\mathbf Z,\bolds\alpha,\bolds\gamma,\bolds
\Pi)}\biggr)
\\[-1pt]
&&\qquad = \sum_{r,s}^{S}
\biggl\{ \biggl(a_{rs}^{0}-a_{rs}+\sum
_{r_i=r,r_j=s}A_{ij} \biggr)\log(\gamma_{rs})
\biggr\}
\\[-1pt]
&&\quad\qquad{}+\sum_{r,s}^{S} \biggl\{
\biggl(b_{rs}^{0}-b_{rs}+\sum
_{r_i=r,r_j=s}(1-A_{ij}) \biggr)\log(1-\gamma_{rs})
\biggr\}
\\[-1pt]
&&\quad\qquad{}+\sum_{s=1}^{S}\sum
_{k=1}^{K} \Biggl\{ \Biggl(\chi_{sk}^{0}-
\chi_{sk}+\sum_{i=1}^{N}
\delta(r_{i}=s)Z_{ik} \Biggr)\log(\alpha_{sk})
\Biggr\}
\\[-1pt]
&&\quad\qquad{}+\sum_{k,l}^{K}\sum
_{c=1}^{C} \Biggl\{ \Biggl(\Xi_{klc}^{0}-
\Xi_{klc}+\sum_{i\neq j}^{N}
\delta(X_{ij}=c)Z_{ik}Z_{jl} \Biggr)\log(
\Pi_{klc}) \Biggr\}
\\[-1pt]
&&\quad\qquad{} - \sum_{i=1}^{N}\sum
_{k=1}^{K}Z_{ik}\log(\tau_{ik})+
\sum_{r,s}^{S}\log\biggl(\frac{B(a_{rs},b_{rs})}{B(a_{rs}^{0},b_{rs}^{0})}
\biggr)
\\[-1pt]
&&\quad\qquad{}+\sum_{s=1}^{S}\log\biggl(
\frac{C(\bolds \chi_{s})}{C(\bolds
\chi_{s}^{0})}\biggr) + \sum_{k,l}^{K}\log
\biggl(\frac{C(\bolds\Xi
_{kl})}{C(\bolds\Xi_{kl}^{0})}\biggr).
\end{eqnarray*}
Therefore, if $\mathcal{L}(q)$ is computed right after the M step,
\begin{eqnarray*}
\mathcal{L}(q)&=&\sum_{r,s}^{S}\log\biggl(
\frac
{B(a_{rs},b_{rs})}{B(a_{rs}^{0},b_{rs}^{0})}\biggr) +\sum_{s=1}^{S}\log
\biggl(\frac{C(\bolds \chi_{s})}{C(\bolds
\chi_{s}^{0})}\biggr)
\\
&&{} + \sum_{k,l}^{K}
\log\biggl(\frac{C(\bolds\Xi
_{kl})}{C(\bolds\Xi_{kl}^{0})}\biggr)-\sum_{i=1}^{N}
\sum_{k=1}^{K}\tau _{ik}\log(
\tau_{ik}).
\end{eqnarray*}\upqed
\end{pf}
\end{appendix}


\begin{supplement}
\stitle{Data and code}
\slink[doi]{10.1214/13-AOAS691SUPP} 
\sdatatype{.zip}
\sfilename{aoas691\_supp.zip}
\sdescription{We provide the original ecclesiastical network along with a
file giving the kingdoms of all vertices in the network and an R code
implementing the variational inference approach for the RSM model.}
\end{supplement}



%

\printaddresses

\end{document}